\documentclass[3p]{elsarticle}

\usepackage[breaklinks=true]{hyperref}
\usepackage[anythingbreaks]{breakurl}
\usepackage{amsthm}
\usepackage{amsmath}
\usepackage{amsfonts}
\usepackage{amssymb}
\usepackage{amsfonts}
\usepackage{epsfig}
\usepackage{bbm}
\usepackage{todonotes}
\usepackage{soul}
\usepackage{nicefrac}
\usepackage{booktabs}
\usepackage{siunitx}

\usepackage{caption}
\usepackage{subcaption}

\newcommand{\tens}[1]{\boldsymbol{\mathrm{#1}}}

\newcommand{\ICdev}{\mathrm{I}_{\tens{\bar C}}}
\newcommand{\IICdev}{\mathrm{II}_{\tens{\bar C}}}

\newcommand{\Rasul}[1]{#1}
\newcommand{\Rasultwo}[1]{{#1}}
\newcommand{\Rasulthree}[1]{{#1}}

\DeclareMathOperator{\tr}{tr}

\journal{Journal of \LaTeX\ Templates}
\usepackage{pgfplots}

\pgfplotsset{compat=newest}

\definecolor{rwth}   {RGB}{  0  84 159}
\definecolor{rwth-75}{RGB}{ 64 127 183}
\definecolor{rwth-50}{RGB}{142 186 229}
\definecolor{rwth-25}{RGB}{199 221 242}
\definecolor{rwth-10}{RGB}{232 241 250}

\definecolor{black}   {RGB}{  0   0   0}
\definecolor{black-75}{RGB}{100 101 103}
\definecolor{black-50}{RGB}{156 158 159}
\definecolor{black-25}{RGB}{207 209 210}
\definecolor{black-10}{RGB}{236 237 237}

\definecolor{magenta}   {RGB}{227   0 102}
\definecolor{magenta-75}{RGB}{233  96 136}
\definecolor{magenta-50}{RGB}{241 158 177}
\definecolor{magenta-25}{RGB}{249 210 218}
\definecolor{magenta-10}{RGB}{253 238 240}

\definecolor{yellow}   {RGB}{255 237   0}
\definecolor{yellow-75}{RGB}{255 240  85}
\definecolor{yellow-50}{RGB}{255 245 155}
\definecolor{yellow-25}{RGB}{255 250 209}
\definecolor{yellow-10}{RGB}{255 253 238}

\definecolor{petrol}   {RGB}{  0  97 101}
\definecolor{petrol-75}{RGB}{ 45 127 131}
\definecolor{petrol-50}{RGB}{125 164 167}
\definecolor{petrol-25}{RGB}{191 208 209}
\definecolor{petrol-10}{RGB}{230 236 236}

\definecolor{turkis}   {RGB}{  0 152 161}
\definecolor{turkis-75}{RGB}{  0 177 183}
\definecolor{turkis-50}{RGB}{137 204 207}
\definecolor{turkis-25}{RGB}{202 231 231}
\definecolor{turkis-10}{RGB}{235 246 246}

\definecolor{grun}   {RGB}{ 87 171  39}
\definecolor{grun-75}{RGB}{141 192  96}
\definecolor{grun-50}{RGB}{184 214 152}
\definecolor{grun-25}{RGB}{221 235 206}
\definecolor{grun-10}{RGB}{242 247 236}

\definecolor{maigrun}   {RGB}{189 205   0}
\definecolor{maigrun-75}{RGB}{208 217  92}
\definecolor{maigrun-50}{RGB}{224 230 154}
\definecolor{maigrun-25}{RGB}{240 243 208}
\definecolor{maigrun-10}{RGB}{249 250 237}

\definecolor{orange}   {RGB}{246 168   0}
\definecolor{orange-75}{RGB}{250 190  80}
\definecolor{orange-50}{RGB}{253 212 143}
\definecolor{orange-25}{RGB}{254 234 201}
\definecolor{orange-10}{RGB}{255 247 234}

\definecolor{rot}   {RGB}{204   7  30}
\definecolor{rot-75}{RGB}{216  92  65}
\definecolor{rot-50}{RGB}{230 150 121}
\definecolor{rot-25}{RGB}{243 205 187}
\definecolor{rot-10}{RGB}{250 235 227}

\definecolor{bordeaux}   {RGB}{161  16  53}
\definecolor{bordeaux-75}{RGB}{182  82  86}
\definecolor{bordeaux-50}{RGB}{205 139 135}
\definecolor{bordeaux-25}{RGB}{229 197 192}
\definecolor{bordeaux-10}{RGB}{245 232 229}

\definecolor{violett}   {RGB}{ 97  33  88}
\definecolor{violett-75}{RGB}{131  78 117}
\definecolor{violett-50}{RGB}{168 133 158}
\definecolor{violett-25}{RGB}{210 192 205}
\definecolor{violett-10}{RGB}{237 229 234}

\definecolor{lila}   {RGB}{122 111 172}
\definecolor{lila-75}{RGB}{155 145 193}
\definecolor{lila-50}{RGB}{188 181 215}
\definecolor{lila-25}{RGB}{222 218 235}
\definecolor{lila-10}{RGB}{242 240 247}

\pgfplotsset{ 
	cycle list/.define={my marks}{ 
		every mark/.append style={solid,fill=\pgfkeysvalueof{/pgfplots/mark list fill}},mark=*\\ 
		every mark/.append style={solid,fill=\pgfkeysvalueof{/pgfplots/mark list fill}},mark=star\\ 
		every mark/.append style={solid,fill=\pgfkeysvalueof{/pgfplots/mark list fill}},mark=square\\ 
		every mark/.append style={solid,fill=\pgfkeysvalueof{/pgfplots/mark list fill}},mark=+\\   
		every mark/.append style={solid,fill=\pgfkeysvalueof{/pgfplots/mark list fill}},mark=triangle\\ 
		every mark/.append style={solid,fill=\pgfkeysvalueof{/pgfplots/mark list fill}},mark=x\\   
		every mark/.append style={solid,fill=\pgfkeysvalueof{/pgfplots/mark list fill}},mark=pentagon\\ 
		every mark/.append style={solid,fill=\pgfkeysvalueof{/pgfplots/mark list fill}},mark=o\\ 
		every mark/.append style={solid,fill=\pgfkeysvalueof{/pgfplots/mark list fill}},mark=diamond\\ 
	}, 
	cycle list/.define={my colors}{rwth, petrol, turkis, grun, maigrun, yellow, orange, magenta, rot, bordeaux, violett, lila},
	cycle multiindex* list={
		my marks\nextlist 
		my colors\nextlist
	}
}

\graphicspath{{./StandalonePics/}}

\begin{document}

\begin{frontmatter}
	\title{Automatic generation of interpretable hyperelastic material models by symbolic regression}
	
	\author[km]{Rasul Abdusalamov\corref{mycorrespondingauthor}}
		\cortext[mycorrespondingauthor]{Corresponding author}
		\ead{abdusalamov@km.rwth-aachen.de}
	\author[km]{Markus Hillg\"artner}
	\author[km]{Mikhail Itskov}

	\address[km]{Department of Continuum Mechanics, RWTH Aachen University, Germany}

	\begin{abstract}		
		In this paper, we present a new procedure to automatically generate interpretable hyperelastic material models. This approach is based on symbolic regression which represents an evolutionary algorithm searching for a mathematical model in the form of an algebraic expression. This results in a relatively simple model with good agreement to experimental data. By expressing the strain energy function in terms of its invariants or other parameters, it is possible to interpret the resulting algebraic formulation in a physical context. In addition, a direct implementation of the obtained algebraic equation is possible.
		
		For the validation of the proposed approach, benchmark tests on the basis of the generalized Mooney-Rivlin model are presented. In all these tests, the chosen ansatz can find the predefined models. Additionally, this method is applied for the multi-axial loading data set of vulcanized rubber. Finally, a data set for a temperature-dependent thermoplastic polyester elastomer is evaluated. In latter cases, good agreement with the experimental data is obtained.		
	\end{abstract}

	\begin{keyword}
		Multi-axial constitutive modeling \sep Hyperelasticity \sep Machine learning \sep Symbolic regression
	\end{keyword}
\end{frontmatter}


\section{Introduction}\label{sec:intro}
Over the past decades, advances in data processing, image acquisition techniques, and fast high-precision sensors significantly increased the amount of data that can be obtained for any engineering task. The question of how to adequately process and incorporate the increasing amounts of data in materials science has gained significant attention in recent years. Most data-driven approaches incorporate machine learning techniques that provide some sort of physical relation obtained directly from experimental data. In context of material modeling, constitutive equations, which provide a physical relation between mechanical stress and strain, are of particular interest.

One type of data-driven constitutive approaches aims to describe the constitutive behavior without an a priori assumption of its functional form. This can be addressed from a pure mathematical point of view \cite{Eggersmann2019,Ibanez2017,Ibanez2018,Kirchdoerfer2016,Leygue2018} or by utilizing artificial neural networks (ANNs) \cite{Hashash2004}. Since approaches of that kind mostly rely solely on stress-strain data, theoretical knowledge from materials theory or thermodynamics is not incorporated, which increases the required amount of data points. Introducing basic concepts of continuum mechanics can help to overcome this issue, which leads to invariant-based ANNs \cite{Linka2020,Shen2004,Liang2008}.
 
Another branch of data-driven constitutive models use representative volume elements (RVEs) to reproduce the microstructure of the material. 
By means of computational homogenization of RVEs, the macroscopic constitutive behavior of the material can be described \cite{Fritzen2013a,Miehe1999}. \Rasul{Approaches based on RVEs are also able to predict the effects of changes in the microstructure, for example due to different grain size distributions.} However, the computational cost can be very high even if accelerated by model order reduction techniques \cite{Fritzen2013a} or machine learning \cite{Le2015,Liu2019,Reimann2019}.

The approaches proposed in the literature differ in the extent to which a physically motivated framework is used as a foundation. For materials whose stress-strain relationship can be derived from a strain energy function $\Psi$ depending on strain invariants, invariant-based artificial neural networks have been developed \cite{Linka2020,Shen2004,Liang2008}. The primary architectural philosophy is here to decouple well-known fundamentals of continuum mechanics from material related aspects. This is advantageous since solely finding a suitable scalar strain energy density function is simpler than that of stress-strain relation. Furthermore, the continuum mechanical framework decreases the likelihood of nonphysical results. Nevertheless, the majority of data-driven approaches share some of the following disadvantages:
\begin{enumerate}
	\item The obtained model represents a ''black box'', thus an interpretation of the results or a critical discussion is not possible.
	\item The model predicts only one special loading case (e.g., uniaxial tension) and is incapable of reasonably describing the behavior when other deformation states are considered.
	\item The obtained model is of large complexity (e.g., because it is given as an artificial neural network with a large number of parameters). This makes  its further applications (e.g., within finite element simulations) cumbersome, inefficient, or even impossible.
	\item If not constrained by appropriate physical conditions, the obtained mathematical equations can potentially be nonphysical (e.g., when they violate Noll's axioms \cite{Truesdell2004}). 
\end{enumerate}

In the present article, a novel invariant-based approach for machine learned data-driven models based on symbolic regression is proposed. This specific type of regression aims to find a closed-form mathematical expression for the strain energy function without assuming any model structure (e.g., linear) a priori. Based on a provided set of allowed input variables, mathematical operators, and basic mathematical functions (e.g., sinus or the exponential function), symbolic regression constructs candidates for further evaluation based on genetic programming.

In genetic programming, inspired by the Darwinian theory of evolution, a starting population of (usually random) mathematical expressions is formed. Utilizing evolutionary concepts, the population evolves over a defined number of generations aiming to become more suitable for the considered problem. In every new generation, some expressions are removed and new once are included. Based on a fitness criterion that defines the quality of the expression for the desired application, expressions are either chosen for deletion or used as a basis to form a slightly modified new one that gets added to the population. In this way, the fitness of the entire population (and especially the fittest member within the population) is assumed to improve over time. 

Multiple applications of symbolic regression in the field of material modeling have been proposed in recent years. The technique can be used in the field of constitutive modeling in combination with a user-chosen material model where symbolic regression is solely responsible to find an expression for each model parameter when the parameters are not constant (e.g., when they depend on the temperature) \cite{Kabliman2021,Kronberger2021,Kronberger2022}. While this hybrid application provides relatively simple solutions, it is unable to utilize symbolic regression to its full potential as the quality of the outcome is still strictly limited by the choice of the base material model.

Kabliman et al. applied symbolic regression to obtain the entire stress-strain curve for one-dimensional loading of aluminum alloys \cite{Kabliman2021,Kabliman2019}. Gusel and Brezocnik made use of genetic programming in order to compute the impact toughness, the tensile strength, and the electrical conductivity of a cold formed material \cite{Gusel2011,Gusel2006}. Asadzadeh et al. applied symbolic regression to the process of sheet bending to investigate the underlying stamping force \cite{Asadzadeh2021}. Wang et al. proposed a tensorial sparse symbolic regression method for determining a relationship between stresses and strains \cite{Wang2022}.

In the field of strain energy based models, symbolic regression has not yet been extensively studied although it has the potential to provide high quality solutions. The focus of this article is to investigate the applicability of symbolic regression and genetic programming with specific attention to this class of materials. A scalar strain energy function found in this way, when utilized within a continuum mechanical framework, is able to provide a constitutive model that can predict arbitrary deformation states. The proposed method is able to formulate strain energy density functions while being fitted against test data obtained from multi-axial loading conditions. Furthermore, it is shown that additional input arguments (such as temperature or other physical parameters) can be incorporated when mechanically relevant.

The article is organized as follows: \autoref{sec:CM} introduces the utilized continuum mechanical framework. The basic principles of symbolic regression and genetic programming are discussed in  \autoref{sec:SRandGP}. The approach implemented in \autoref{sec:implementation} is validated against multiple sets of artificially generated and experimentally obtained data sets in \autoref{sec:results}. Finally, a brief conclusion highlighting the main aspects of this work is presented in \autoref{sec:conclusion}.


\section{Continuum mechanical framework}\label{sec:CM}
Constitutive models that derive a stress–strain relationship from a strain energy density function $\Psi$ are commonly applied for biological soft tissues (e.g., \cite{Chen2015,Gasser2006,Hillgartner2019a,Linka2018,Rynkevic2018}) and rubber-like materials (e.g., \cite{Linka2020,Arruda1993,Boyce2000,Kaliske1999,Pham2021}). In the following, we assume the strain energy function $\Psi$ of a nearly-incompressible material to be function of the invariants of the right Cauchy-Green tensor $\tens{C}$.

Note that the approach presented in this article does not depend on a particular choice of arguments of the strain energy function. Applications incorporating various anisotropy classes (e.g., with a larger invariant set as considered in \cite{Linka2020,Ehret2007,Itskov2004b}), material characteristics (e.g., filler content or fiber diameter), or additional relevant information (e.g., temperature or humidity) can be considered in a straight-forward fashion.

The principal invariants of the right Cauchy-Green tensor  $\tens{C}$ can be expressed as 
\begin{align}
	\label{equ:C_invariants}
	\mathrm{I}_{\tens{C}} &= \tr\tens{C} \ ,
	&
	\mathrm{II}_{\tens{C}} &= \dfrac{1}{2}\left[(\tr\tens{C})^2 -\tr\left(\tens{C}^2\right)\right] \ ,	
	&
	\mathrm{III}_{\tens{C}} &= \det\tens{C} \ .
\end{align}
Many materials of practical interest, especially rubbers or materials with high water content like biological tissues, are characterized by nearly incompressible behavior. The incompressibility can be taken into account by the constraint $J=\det\tens{F}=1$. A  numerically more convenient approach widely used in computational mechanics is to allow small volume changes, considering the material as nearly-incompressible. This can be achieved by a multiplicative split of the deformation gradient $\tens{F}$ into an isochoric (volume-preserving) part $\tens{\bar F}$ and a volumetric (volume changing) part $\tens{\hat F}$ as \cite{Flory1961}
\begin{align*}
	\tens{F} &= \tens{\hat F} \tens{\bar F}\ ,
	&
	\tens{\hat F} &= J^{1/3}\tens{I} \ ,
	&
	\tens{\bar F} &= J^{-1/3}\tens{F} \ ,
	&
	\det \tens{\bar F} &= 1 \ .
\end{align*}
Thus, for the isochoric right Cauchy-Green tensor $\tens{\bar C}$ holds
\begin{equation}
	\tens{\bar C} = \tens{\bar F}^\mathrm{T}\tens{\bar F} = J^{-2/3}\tens{C} \ .
\end{equation}
Similarly, it is possible to determine the principal invariants of $\tens{\bar C}$ as
\begin{align}
	\label{equ:Cbar_invariants}
	\mathrm{I}_{\tens{\bar C}} &= J^{-2/3}\mathrm{I}_{\tens{ C}} \ ,
	&
	\mathrm{II}_{\tens{\bar C}} &= J^{-4/3}\mathrm{II}_{\tens{ C}} \ ,
	&
	\mathrm{III}_{\tens{\bar C}} = 1 \ .
\end{align}
For a strain energy function depending on the invariants \eqref{equ:Cbar_invariants}
\begin{equation}
	\label{equ:Psi2}
	\bar\Psi = \bar\Psi(\mathrm{I}_{\tens{\bar C}},\mathrm{II}_{\tens{\bar C}},J),
\end{equation}
the first Piola-Kirchhoff stress tensor $\tens{P}$ can be obtained as 
\begin{equation}
	\label{equ:P2}
	\tens{P}
	= \dfrac{\partial \bar\Psi}{\partial \tens{F}}
	= 2\left(\dfrac{\partial\bar\Psi}{\partial\mathrm{I}_{\tens{\bar C}}}  J^{-2/3} \tens{F}
		+ \dfrac{\partial\bar\Psi}{\partial\mathrm{II}_{\tens{\bar C}}}  J^{-2/3} \left(\mathrm{I}_{\tens{\bar C}}\tens{F} - J^{-2/3}\tens{F}\tens{C}\right)
		+ \dfrac{\partial\bar\Psi}{\partial J} \dfrac{J}{2} \tens{F}^\mathrm{-T}\right).
\end{equation}
Hence, in order to obtain the stress, the derivatives of the strain energy density function $\bar\Psi$ with respect to the invariants $\mathrm{I}_{\tens{\bar C}}$, $\mathrm{II}_{\tens{\bar C}}$, and $J$ have to be determined.


\section{Symbolic regression and genetic programming}\label{sec:SRandGP}
In statistical modeling, regression analysis refers to methods that aim to determine mathematical relations between one or more independent variables and a single dependent variable. Most forms of regression are provided with a particular kind of a starting model which needs to be parameterized optimally. For example, the commonly applied linear regression assumes an underlying linear relationship. While this approach can be beneficial when domain-specific knowledge is available, it can face negative effects due to human bias or unknown gaps in the users' understanding of physical relations that need to be described.

Symbolic regression (SR) \cite{Koza1994,Koza1997}, is a special type of regression analysis that does not rely on a user-chosen starting model. In contrast, a large set of mathematical expressions is searched to find the most suitable one. It is hence mostly independent of human bias or a lack of domain knowledge. 

In the following, a brief introduction of genetic programming for the use in symbolic regression is provided. For further detail, the interested reader is referred, among others, to \cite{Koza1994,Wang2019a,Augusto2000}.

\begin{figure}[h]
	\includegraphics[width=\textwidth]{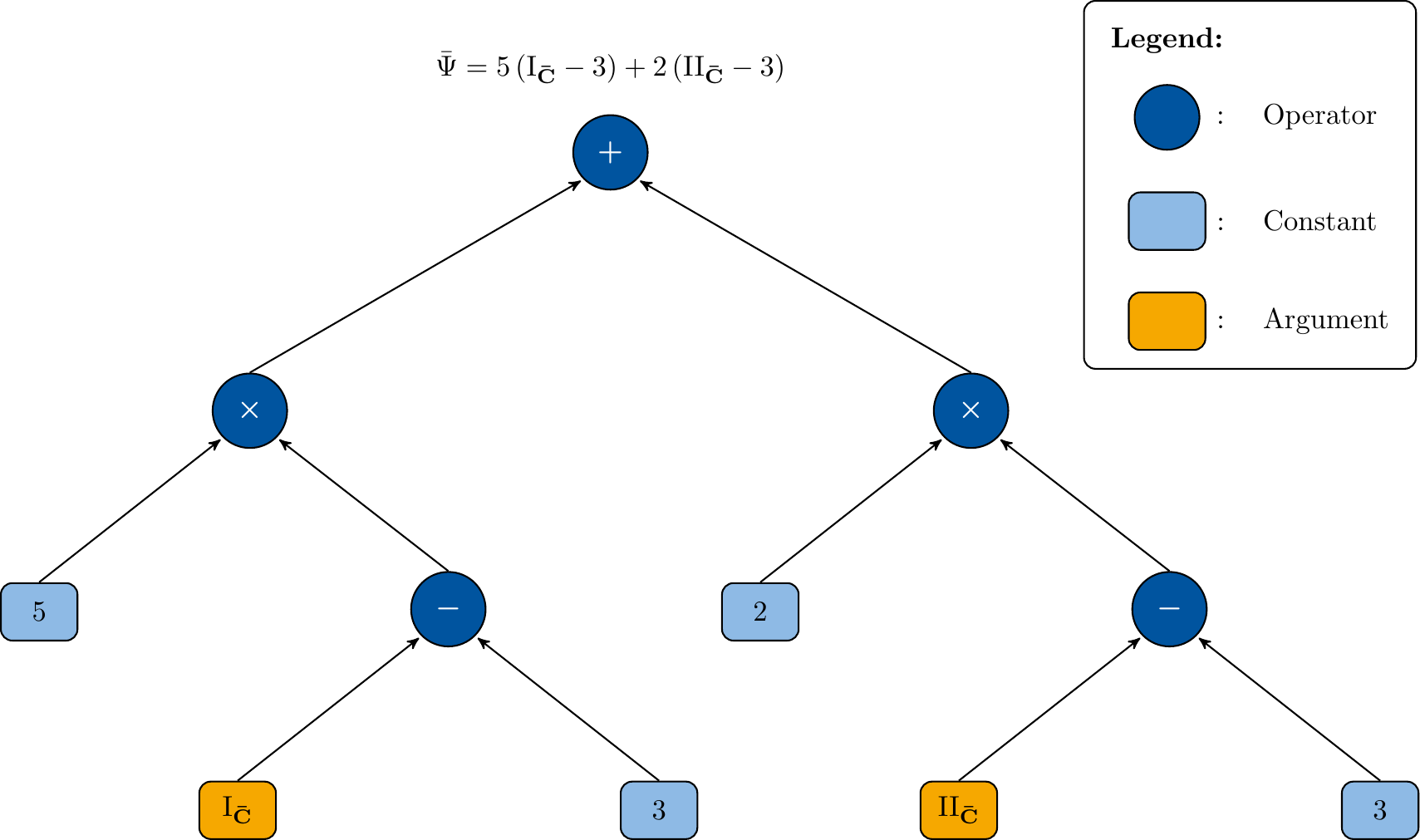}
	\caption{Example of a calculation tree for a strain energy function according to \eqref{equ:Psi2}.}
	\label{fig:compGraph_neoHookean}
\end{figure}

Mathematical expressions, such as strain energy functions, can be represented graphically as a rooted tree. These trees, often referred to as calculation trees (or computational graphs), are traversed beginning in the root. For every vertex within the tree, the left-hand side children are evaluated before the vertex itself followed by the children on the right-hand side. An example of a calculation tree is illustrated in \autoref{fig:compGraph_neoHookean}. The reading order for the left half of the tree is consequently $5$, $\times$, $\mathrm{I}_{\tens{\bar C}}$, $-$, and $3$, relating to the mathematical expression $\left(5\times\left(\mathrm{I}_{\tens{\bar C}}-3\right)\right)$. Vertices which are not succeeded are called external. They contain either one of the independent values (inputs) or constants. Vertices which are succeeded are referred to as internal. They represent mathematical operators or basic functions that connect its following vertices.

\subsection{Initialization}
\Rasul{In the initialization phase, the genetic algorithm generates a population set of random mathematical expressions (represented as computational graphs) based on a user-defined list of allowed basic operators, mathematical functions, the predefined independent variables, and numerical constants. The initial population (or the first generation) of expressions has been found. Due to the randomness, the chosen equations will most likely perform poorly for the designated purpose.}	

The following selection and evolution phases are repeated for a user-defined number of generations.

\subsection{Selection}
Based on the previous generation of mathematical expressions, the next generation needs to be formed. In order to decide which of the expressions within the previous population get to evolve into the succeeding one, tournaments can be performed. All members of the previous generation are grouped into random subsets in which they compete against each other. For every subgroup, the fittest individual (which describes the underlying data best) is selected as basis for the next generation. The tournament size determining how many individuals compete at once, has a significant impact on the speed of the process (since it defines how many individuals are sorted out) and on the diversity of future generations (since more individuals are considered further).

Another method of selection can be fitness proportionate, where the likelihood of the selection of any individual is related to its fitness on the provided data.

\subsection{Evolution}
Multiple possible forms of evolution can be found in the literature for genetic programming. In the following, the most important of them are briefly introduced.

\textit{Crossover} takes the winner of two different tournaments to form a new individual. Within the first winner, referred to as the original parent, a random subtree is selected to be replaced. Similarly, within the second winner, referred to as the donor, a subtree is selected to be the replacement. In the following, the replacement from the donor is inserted into the original parent to create a new individual for the next generation. Mathematically, this is equivalent to replacing a random term within one expression by a random term from the second expression.

\textit{Subtree mutation} is similar to crossover, but only requires the original parent. The missing replacement is randomly generated. Mathematically, this is equivalent to replacing a random term within the expression by a random new term. The subtree mutation allows the reintroduction of operators and mathematical functions that may have died out previously and hence increases the diversity of the population.

\textit{Point mutation} takes the winner of a tournament and replaces random vertices with similiar ones (i.e., functions are replaced by functions with the same number of arguments, and constants are replaced by constants). Mathematically, this is equivalent to replacing a single random operator, function, input, or constant with another one. The point mutation allows the reintroduction of operators and mathematical functions that may have died out previously and hence increases the diversity of the population.

\textit{Hoist mutation} aims to slim existing solutions by replacing a random subtree with a subtree of itself. Mathematically, this is equivalent to replacing a random term with one of its subterms and hence shortening the overall equation.

Just like in nature, the combination of fitness-based selection criteria (for the choice of individuals for the next round) and evolution strategies (that randomly alter or mix individuals) is expected to increase the overall fitness of the entire population over time. While the procedure guarantees syntactically correct mathematical expressions as output, it neither results in the same deterministic solution every time nor guarantees to find the optimal solution.


\section{Implementation}\label{sec:implementation}
In order to obtain a tool able to combine the constitutive framework, described in \autoref{sec:CM}, with the concepts of symbolic regression and genetic programming, as discussed in \autoref{sec:SRandGP}, the relevant continuum mechanics equations need to be incorporated. Available toolboxes for genetic programming are designed to directly find a mathematical relation between the inputs and output.

\Rasul{In the continuum mechanical case, the inputs (e.g., the components of the deformation gradient) are not the direct arguments of the first Piola-Kirchhoff stress tensor. For the fitting, the strain energy density function must be found based on the derivatives with respect to the invariants.} Furthermore, the found strain energy density function cannot be experimentally obtained and hence only a stress value, resulting from differentiation, can be utilized for evaluation of the model performance.

\Rasul{The equations provided in \autoref{sec:CM} are generally valid for hyperelastic nearly-incompressible materials depending on the invariants $\mathrm{I}_{\tens{\bar C}}$, $\mathrm{II}_{\tens{\bar C}}$ and $J$. Certainly, the approach can be extended for any other invariants.  Thus, only the strain energy density and its derivatives with respect to the invariants are material-dependent and no material-specific additions need to be made for a symbolic regression toolbox. Using this continuum mechanical framework with the extension by symbolic regression provides some advantages:}
\begin{enumerate}
	\item The strain energy function obtained by symbolic regression automatically satisfies the material objectivity condition since it is formulated in terms of objective invariants.
	\item By choosing the invariants that are used as input arguments, the user can allow only isotropic or anisotropic models within pre-defined symmetry classes.
	\item The scalar-valued strain energy function is easier to interpret than any other data-driven model that directly relates a deformation tensor to a stress tensor.
	\item Finite element codes provide programming interfaces in which material models can be implemented directly by the user. In some cases (e.g., UHYPER subroutine in Abaqus), the strain energy density and its derivatives with respect to the invariants are required. These equations can be obtained with relative ease.
\end{enumerate}

\begin{figure}[h!]
	\centering
	\includegraphics[width=0.8\textwidth]{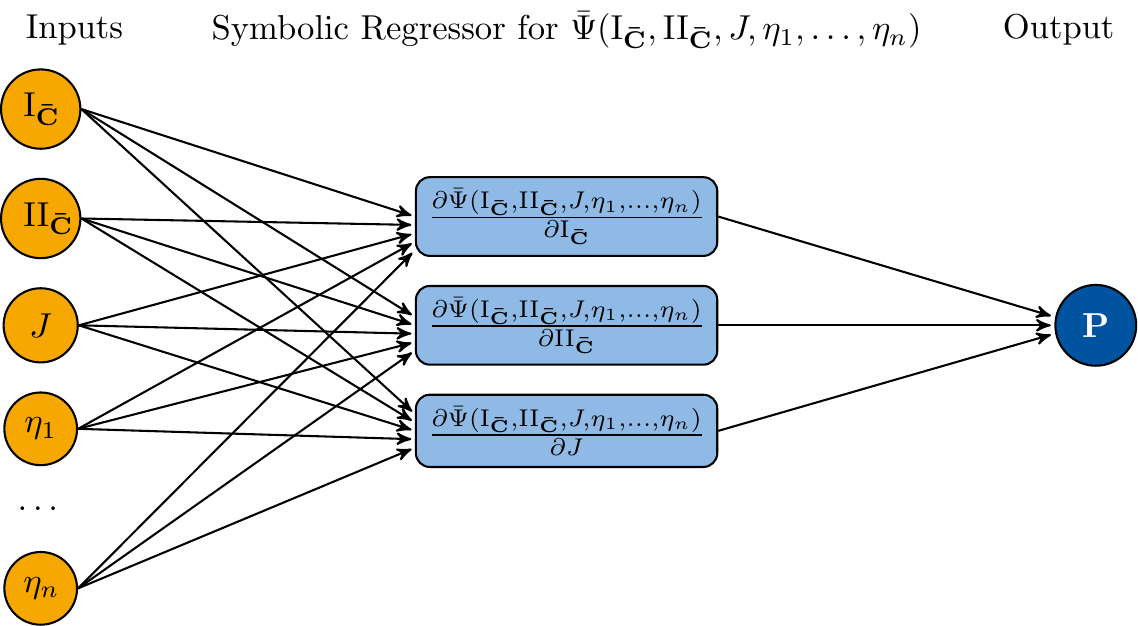}
	\caption{Illustration of the procedure with the invariants $\ICdev$, $\IICdev$, and $J$ as well as further parameters $\eta_1, \, \dots, \, \eta_n$ as inputs and the first Piola-Kirchhoff stress tensor  $\tens{P}$ as the output. The symbolic regressor is responsible for obtaining the strain energy function, where the derivatives are determined with respect to the invariants and evaluated for the fitting.}
	\label{fig:Process}
\end{figure}

Apart from the invariants, strain energy density functions may depend on an arbitrary number of additional parameters, such as the temperature or a specific concentration of material constituents. A general process diagram including invariants and $i=1,\dots,n$ additional parameters $\eta_i$ is visualized in \autoref{fig:Process}.

The results presented in the following have been obtained by utilizing a customized installation of the {\tt scikit-learn}\footnote{\url{https://scikit-learn.org/}} inspired Python package {\tt gplearn}\footnote{\url{https://gplearn.readthedocs.io/}}. Apart from tailoring the inputs and outputs to a desired format and providing the continuum mechanical equations that enclose the strain energy function, no substantial changes have been made. For the sake of simplification, differentiation, as necessary in equation \eqref{equ:P2}, has been performed numerically using difference quotients.

\section{Results}\label{sec:results}
In the following, the presented approach is validated in three different test cases. First, an artificial data set based on a known constitutive model was generated and the capability of the genetic programming to recognize the correct model was assessed as benchmark case. Second, multi-axial loading data of vulcanized rubber reported by Treloar \cite{Treloar1944} was utilized to get a strain energy density function that is able to describe the tested material. Third, the ability to incorporate further information (e.g., the temperature) is demonstrated.

\subsection{Benchmark test}
In order to test the capability of the genetic programming approach to identify reasonable material models based on provided data, artificial data sets generated based on a known constitutive equation can be utilized. Due to the existence and prior knowledge of the ideal solution, the performance of symbolic regression can be easily assessed. Ideally, the genetic algorithm could find the exact solution.

The generated data set is based on a generalized form of the Mooney-Rivlin model \cite{Bower2010,Mooney1940} described by the strain energy density function $\bar\Psi_{\rm gMR}$ as follows
\begin{equation}
	\label{equ:gmR}
	\bar\Psi_{\rm gMR} = 
		\sum\limits_{i=1}^3 \left[c_{i0}\left(\mathrm{I}_{\tens{\bar C}}-3\right)^i + c_{0i}\left(\mathrm{II}_{\tens{\bar C}}-3\right)^i \right] \ ,
\end{equation}
where the parameters were chosen according to \autoref{tab:gMR_par}. 
\begin{table}[h]
	\centering
	\begin{tabular}{ccccccc}
		\toprule
		Case & $c_{10}$ & $c_{20}$ & $c_{30}$ & $c_{01}$ & $c_{02}$ & $c_{03}$ \\
		& $ \left[ \SI{}{\mega \Pa} \right] $ & $ \left[ \SI{}{\mega \Pa} \right] $ & $ \left[ \SI{}{\mega \Pa} \right] $ & $ \left[ \SI{}{\mega \Pa} \right] $ & $ \left[ \SI{}{\mega \Pa} \right] $ & $ \left[ \SI{}{\mega \Pa} \right] $   \\
		\midrule
		1 & 0.48 & 0.00 & 0.00 & 0.12 & 0.00 & 0.00  \\ 
		2 & 0.87 & 0.86 & 0.00 &  0.98 & 0.43 & 0.00  \\ 
		3 & 0.91 & 0.57 & 0.79 & 0.88 & 0.21 & 0.70  \\
		\bottomrule
	\end{tabular}
	\caption{Material parameters used for the generalized Mooney-Rivlin strain energy function as given in equation \eqref{equ:gmR}.}
	\label{tab:gMR_par}
\end{table}
For the three cases given in \autoref{tab:gMR_par}, the material responses under uniaxial tension (UT), pure shear (PS), and equibiaxial tension (EBT) have been calculated. For all cases, a total of five formulations of the strain energy function were sought. The mean value and standard deviation for each strain value of all strain energy functions were determined.

\begin{figure}[h!]
	\begin{subfigure}[b]{0.49\textwidth}
		\includegraphics[width=0.9\textwidth]{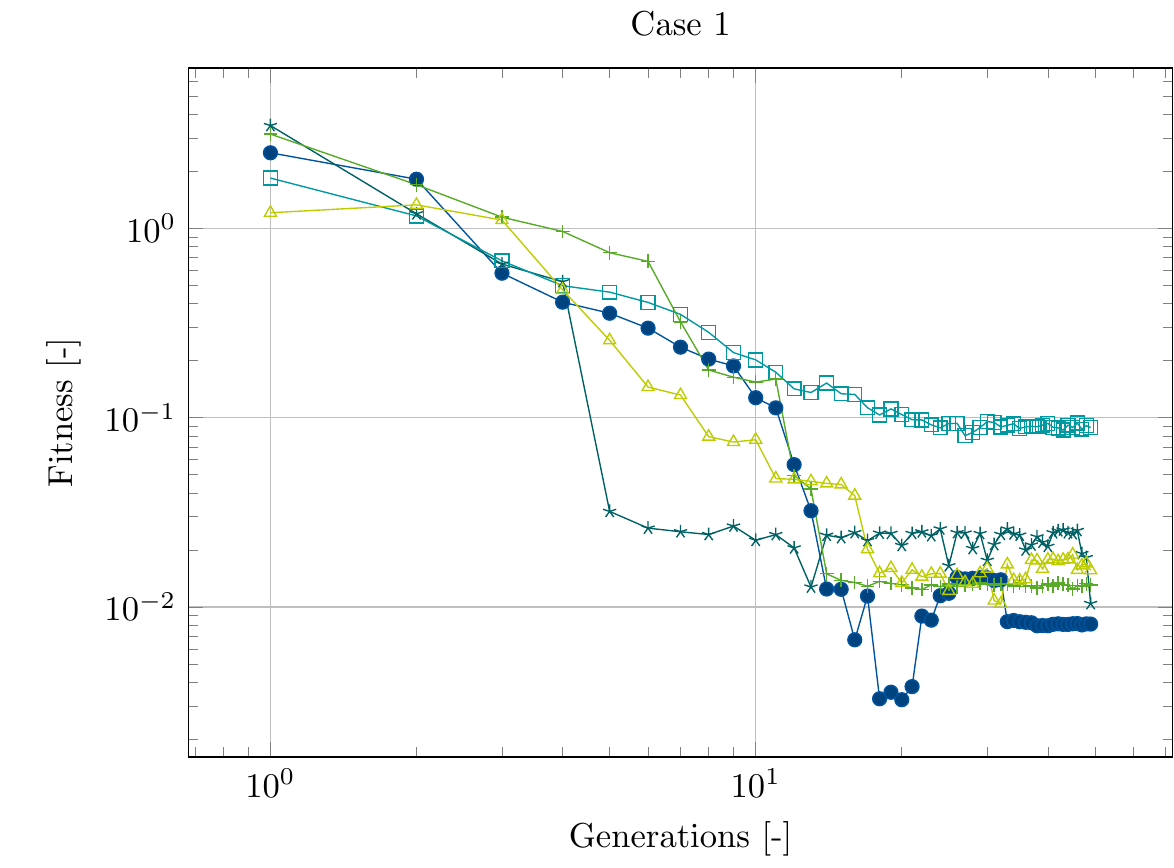}
		\caption{}
		\label{fig:MR_2_f1}
	\end{subfigure}
	\begin{subfigure}[b]{0.49\textwidth}
		\includegraphics[width=0.9\textwidth]{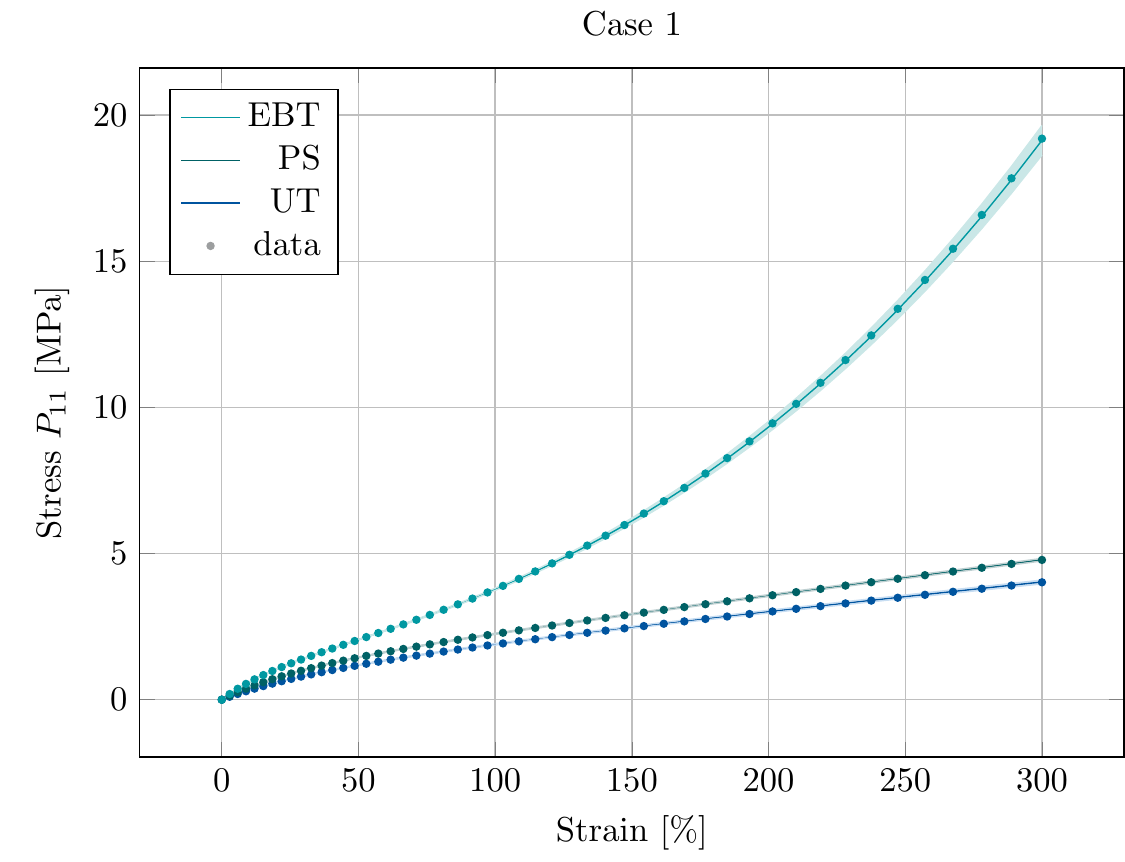}
		\caption{}
		\label{fig:MR_2_f2}
	\end{subfigure}
	\\
	\begin{subfigure}[b]{0.49\textwidth}
		\includegraphics[width=0.9\textwidth]{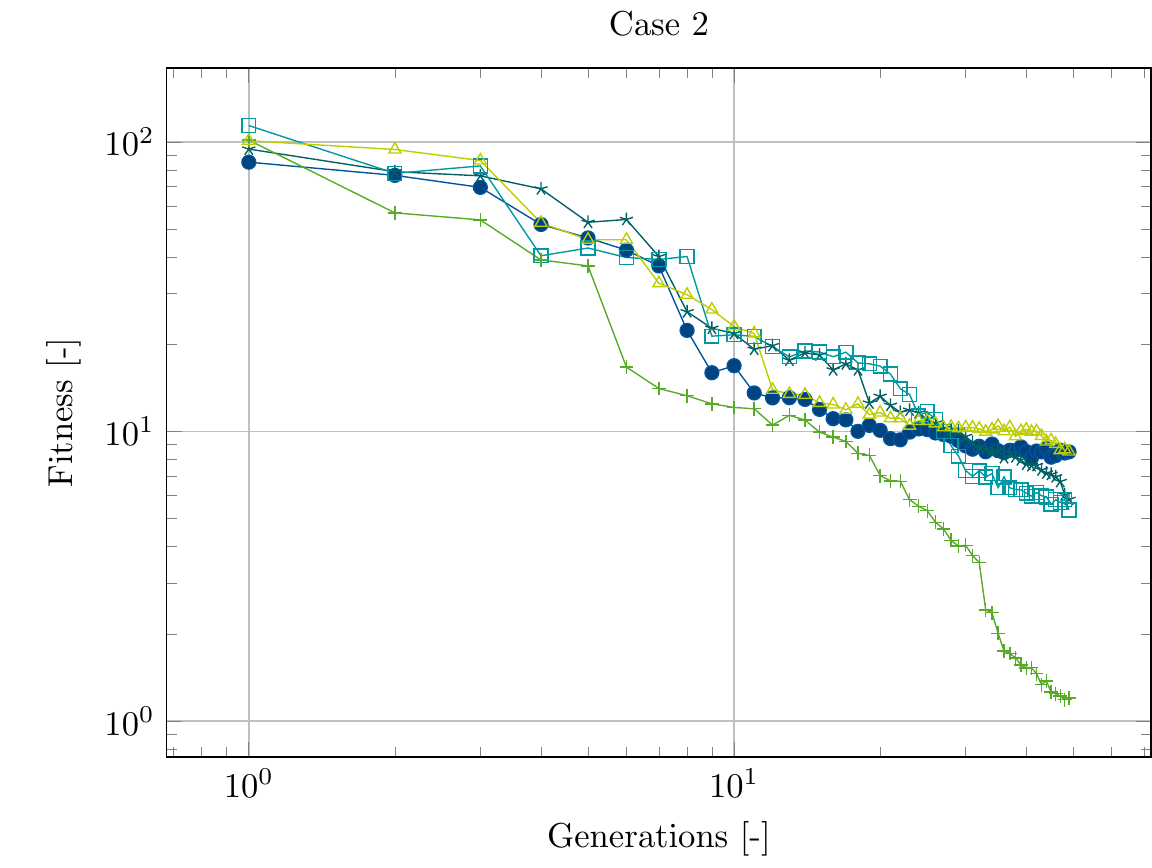}
		\caption{}
		\label{fig:MR_4_f1}
	\end{subfigure}
	\begin{subfigure}[b]{0.49\textwidth}
		\includegraphics[width=0.9\textwidth]{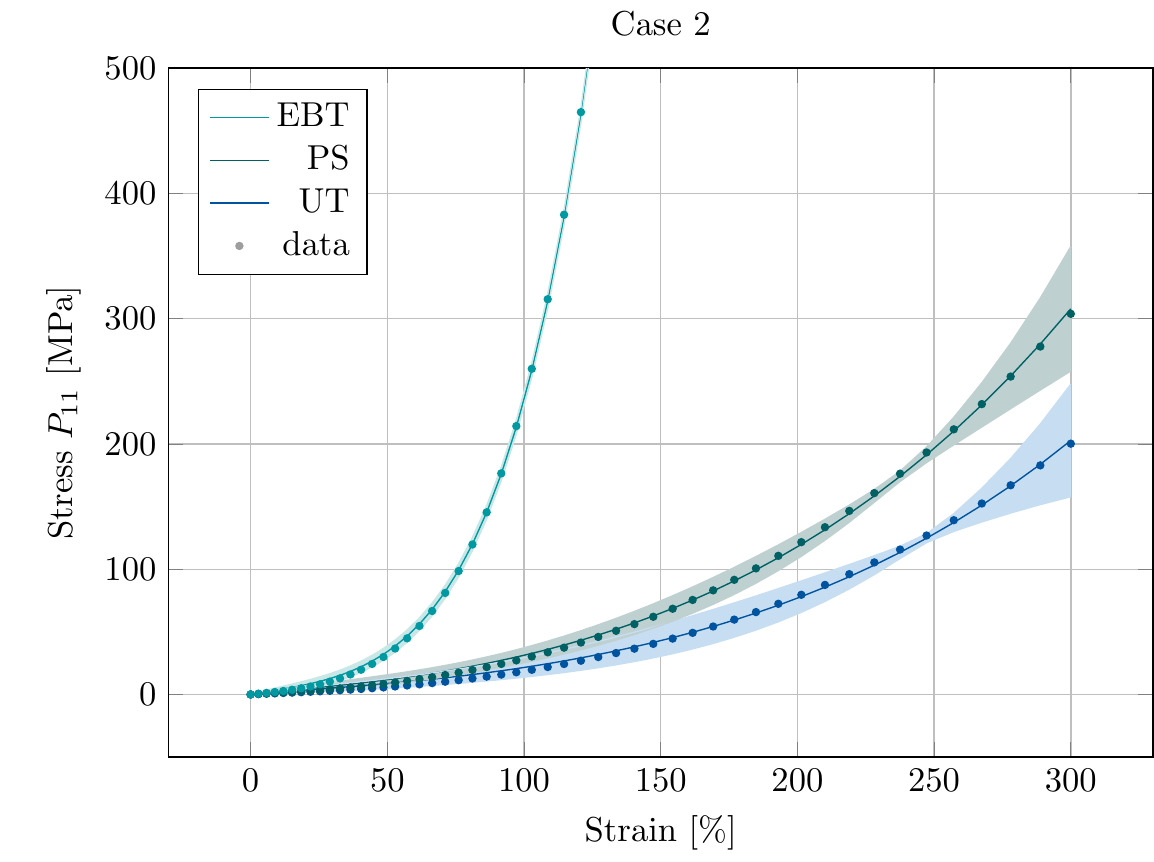}
		\caption{}
		\label{fig:MR_4_f2}
	\end{subfigure}
	\\
	\begin{subfigure}[b]{0.49\textwidth}
		\includegraphics[width=0.9\textwidth]{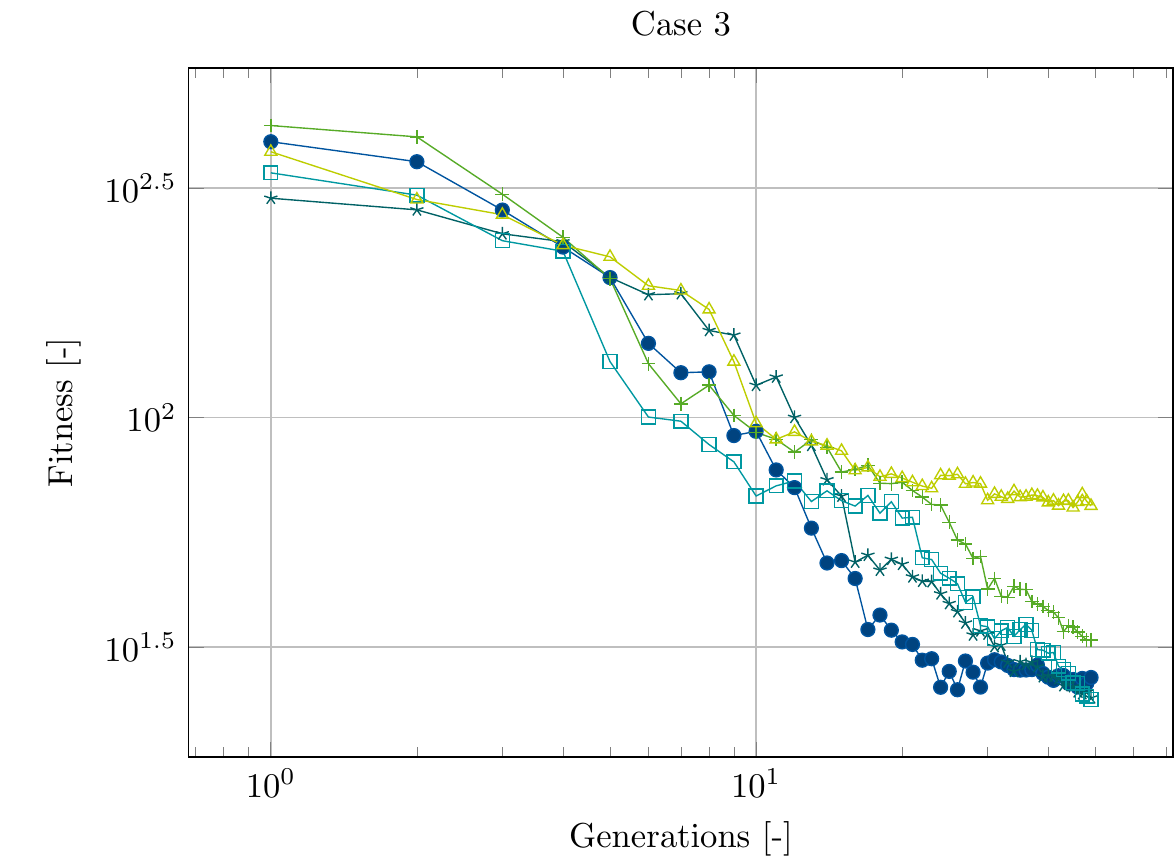}
		\caption{}
		\label{fig:MR_6_f1}
	\end{subfigure}
	\begin{subfigure}[b]{0.49\textwidth}
		\includegraphics[width=0.9\textwidth]{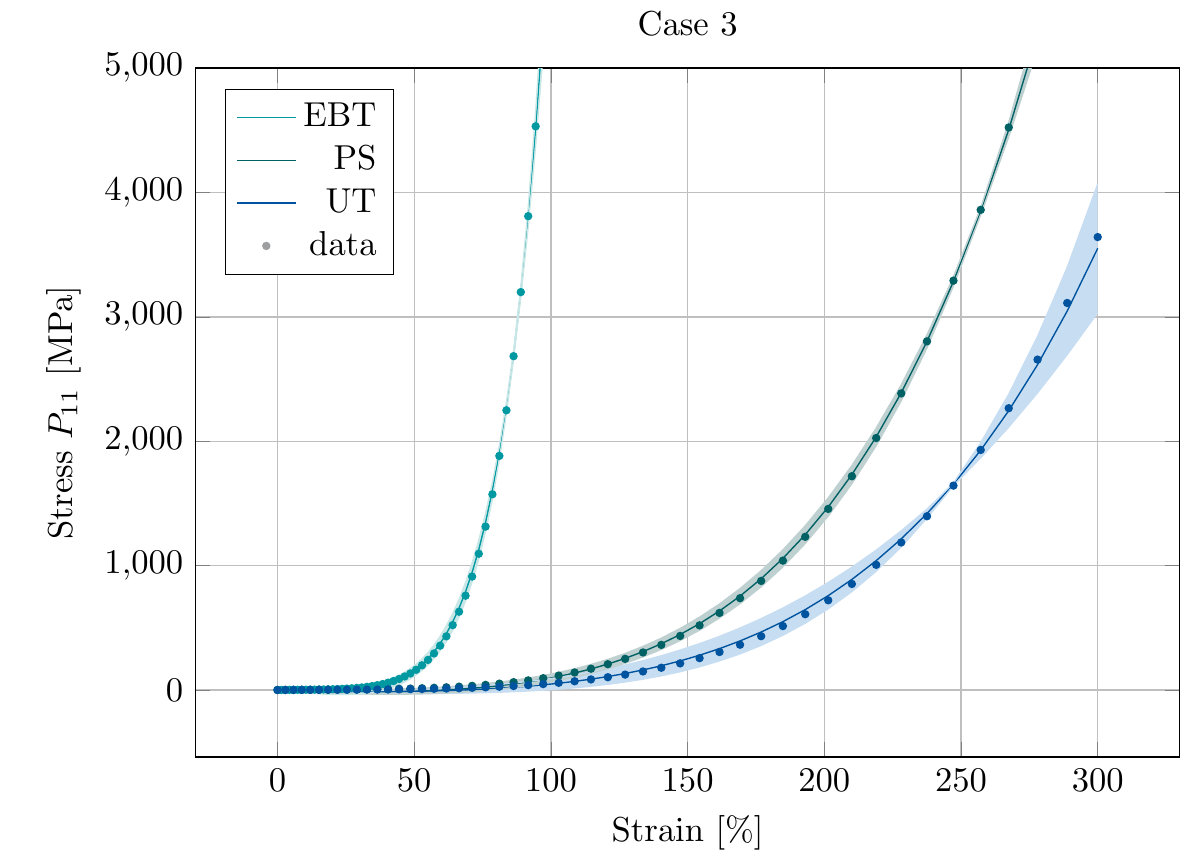}
		\caption{}
		\label{fig:MR_6_f2}
	\end{subfigure}
	\caption{Visualization of the results obtained for five attempts for each of the three Mooney-Rivlin cases provided in \autoref{tab:gMR_par}. \ref{fig:MR_2_f1}, \ref{fig:MR_4_f1}, and \ref{fig:MR_6_f1} represent the fitness over the generations and \ref{fig:MR_2_f2}, \ref{fig:MR_4_f2}, and \ref{fig:MR_6_f2} depict the first Piola-Kirchhoff stress (mean and 6$\sigma$ confidence interval) for uniaxial tension (UT), pure shear (PS), and equibiaxial tension (EBT) loading.}
	\label{fig:MR_incomp_Fitness}
\end{figure}

In \autoref{fig:MR_incomp_Fitness} the fitness is plotted over the generations as well as the first Piola-Kirchhoff stress is visualized as a function of strain. The default fitness in {\tt gplearn} is the mean absolute error (MAE), determined by
\begin{align}
\text{MAE} = \frac{1}{n} \sum_{i=1}^{n} |y_{\text{pred}, i} - y_{i}| \ ,
\end{align}
where $n$ is the number of predicted values, $y_{\text{pred}, i}$ the prediction and $y_{i}$ the true value. For all three cases, the used inputs such as the number of generations or the population sizes are listed in \autoref{tab:ParametersOverview}.

\begin{figure}[h!]
	\begin{subfigure}[b]{0.49\textwidth}
		\includegraphics[width=0.9\textwidth]{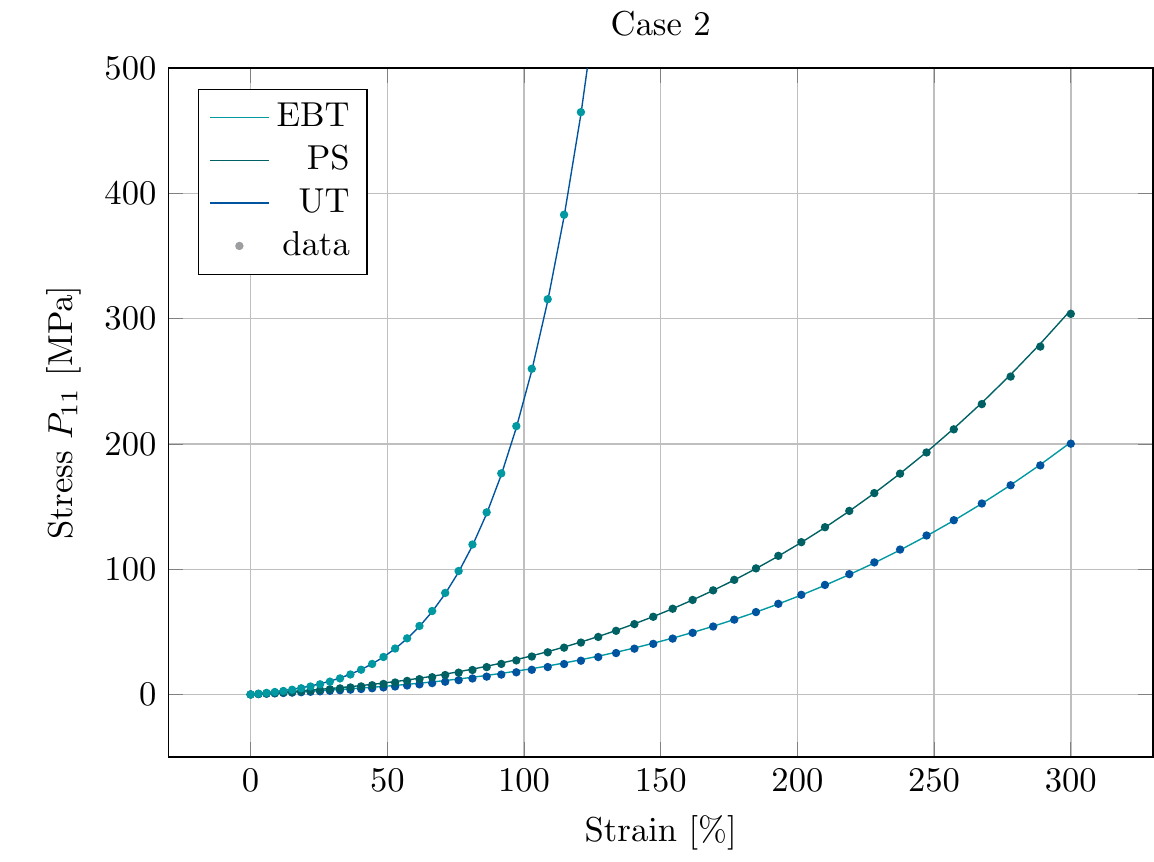}
		\caption{}
		\label{fig:MR_4_BestFit}
	\end{subfigure}
	\hfill
	\begin{subfigure}[b]{0.49\textwidth}
		\includegraphics[width=0.9\textwidth]{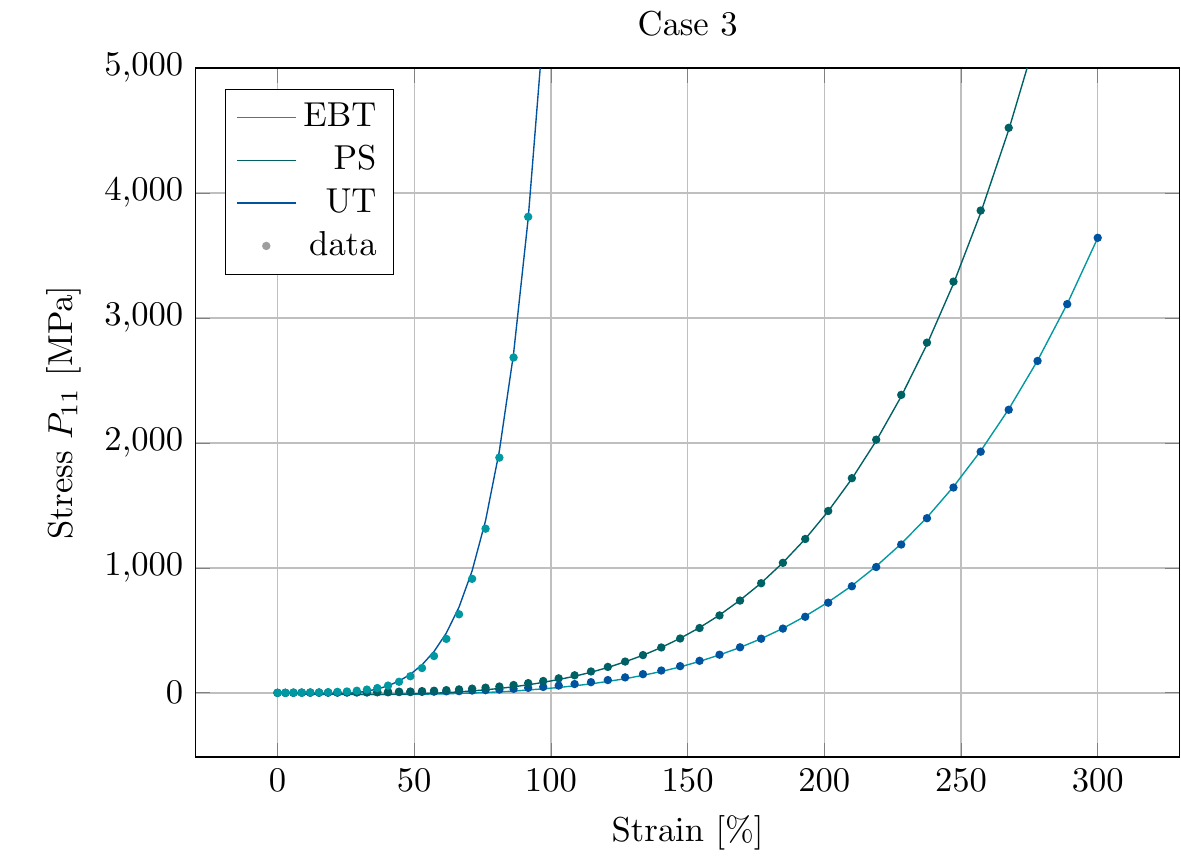}
		\caption{}
		\label{fig:MR_6_BestFit}
	\end{subfigure}
	\caption{Visualization of best fits (out of the generated five) for case 2 and 3.}
	\label{fig:MR_BestFits}
\end{figure}

As the number of generations increases, the result of the fitting improves significantly. Due to the fact that this is an evolutionary algorithm, the best fitting strain energy function formulation is not preserved and may eventually mutate further in the process. Additionally, the average responses for uniaxial tension, pure shear, and equibiaxial tension were determined and the corresponding $6\sigma$ intervals are plotted. In this context, it should be noted that the confidence interval increases with the number of terms. As more terms need to be found, it is less likely that they will be determined quickly. Nevertheless, the averaged predictions coincide almost exactly with the corresponding provided responses. In addition, the best fits for the cases with 4 and 6 terms are presented in \autoref{fig:MR_BestFits}. \Rasultwo{Note that theoretically, a perfect fit can always be found with a very high number of generations and a high population size. The exact formulation of the strain energy function is always retrieved. The constants $c_{i0}$ and $c_{0i}$ can only differ marginally in the decimal digits from the original values.}

\subsection{Multi-axial loading of vulcanized rubber}
\begin{figure}[h!]
\centering
\includegraphics[width=0.5\textwidth]{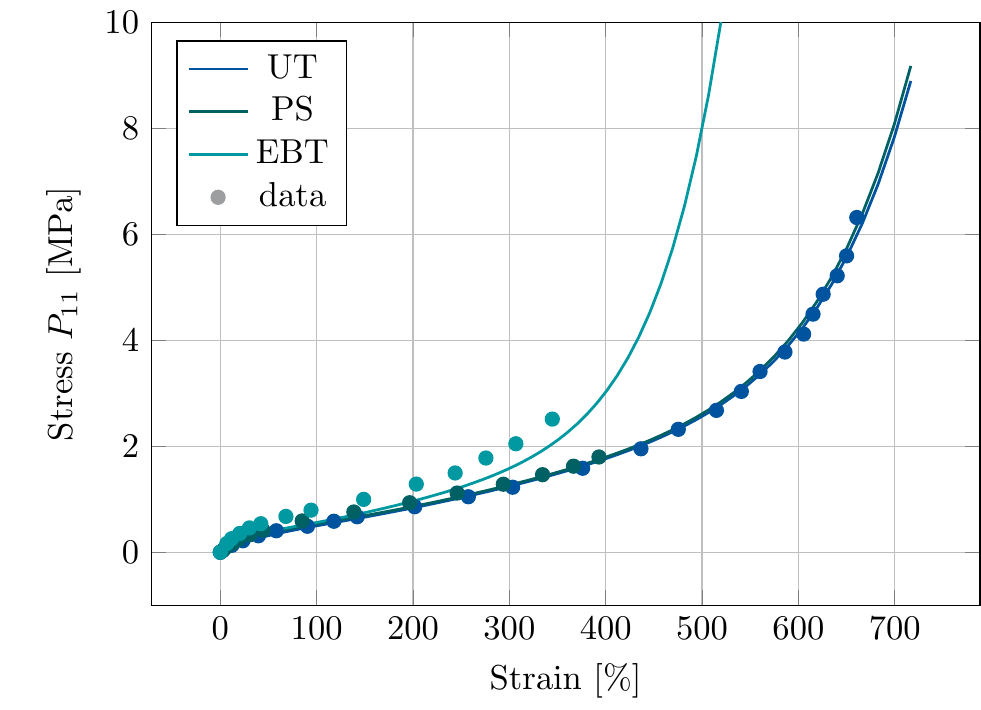}
\caption{Best fit for Treloar data set using the strain energy density function from equation \eqref{equ:TreloarBestFit}.}
\label{fig:TreloarFit}
\end{figure}

In 1944, Treloar \cite{Treloar1944} reported a data set of vulcanized rubber containing stress-strain curves of uniaxial tension (UT), equibiaxial tension (EBT), and pure shear (PS) experiments under large strains. This data set has become one of the best known and most used ones and is still seen as an important benchmark test for constitutive models of rubbers and rubber-like materials \cite{Marckmann2006}. In order to find a suitable material model, it is not enough to fit each of the three curves individually. All three curves need to be considered simultaneously. Based on this data set, a symbolic strain energy density formulation was found as:
\begin{equation}
\label{equ:TreloarBestFit}
\bar\Psi_{\text{Treloar}} = \sqrt{0.93296 \exp\left(0.080316 \mathrm{I}_{\tens{\bar C}}\right) + \sqrt{ \mathrm{I}_{\tens{\bar C}} -0.080316} + \left(0.0232113 \mathrm{I}_{\tens{\bar C}} + 0.021633\right) \mathrm{I}_{\tens{\bar C}}} \ .
\end{equation}
The response of this hyperelastic model is visualized in \autoref{fig:TreloarFit}. The fit was generated with the inputs specified in \autoref{tab:ParametersOverview}. It is important to note that each data point has been given the same weighting for the fitting. The predicted response for uniaxial tension and pure shear is almost perfect, while the equibiaxial tension case is slightly underestimated. Nonetheless, the found material model describes the data set better than many traditional hyperelastic models, such as the Mooney model or the van der Waals model (c.f. \cite{Marckmann2006}). One also observes that the first invariant appears to be sufficient to represent the stress response for all three loading cases.
 
\subsection{Temperature dependency}
\begin{figure}[h!]
	\begin{subfigure}[b]{0.49\textwidth}
		\includegraphics[width=0.95\textwidth]{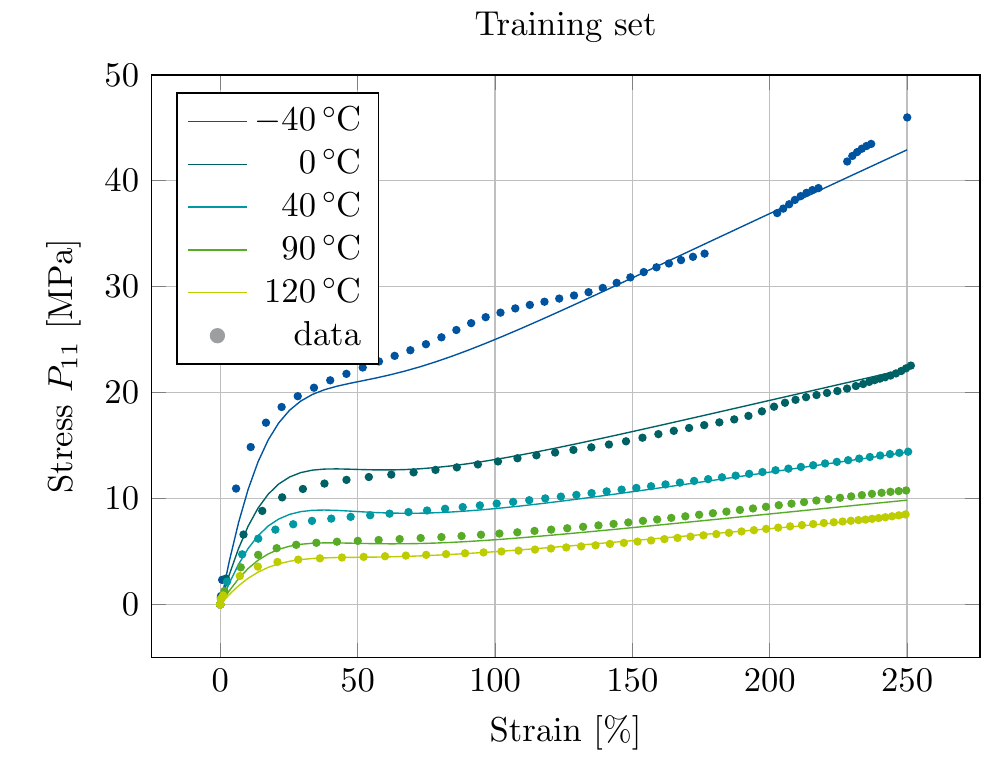}
		\caption{}
		\label{fig:TempFitGivenCurves}
	\end{subfigure}
	\begin{subfigure}[b]{0.49\textwidth}
		\includegraphics[width=0.95\textwidth]{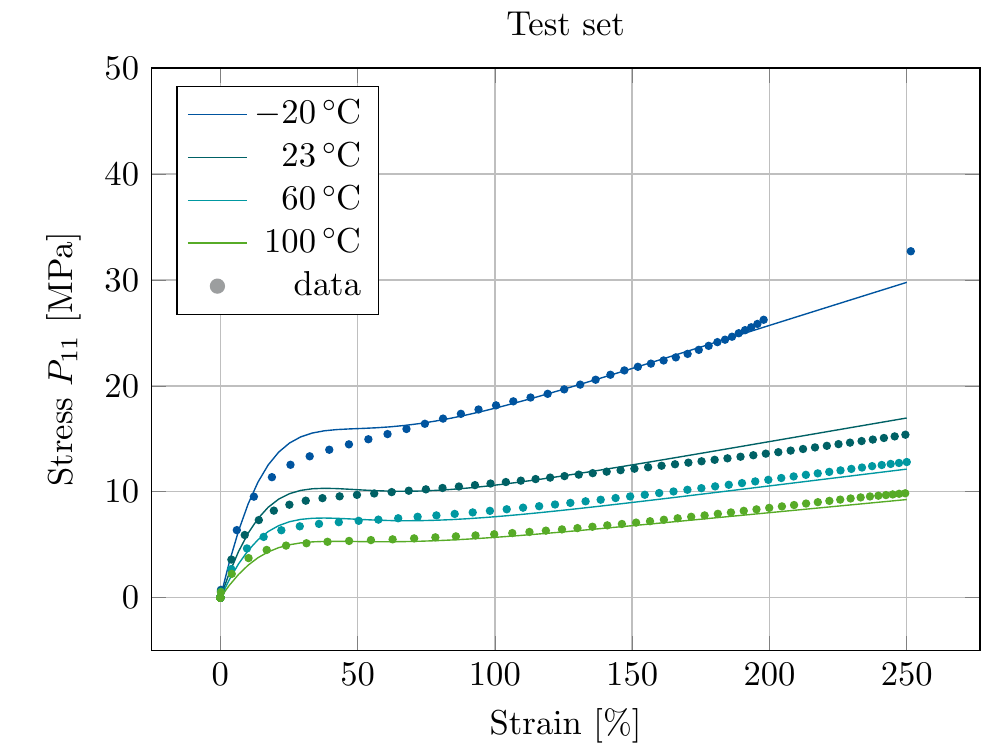}
		\caption{}
		\label{fig:TempFitUnknownCurves}
	\end{subfigure}
	\caption{Temperature dependent stress-strain curves (prediction, equation \eqref{equ:HytrelBestFit}, and reference \cite{Hytrel}) for the training set and the test set.}
	\label{fig:TempFit}
\end{figure}

As discussed earlier in \autoref{sec:implementation}, other parameters can be incorporated as well. Here, a temperature-dependent data set will be considered. We use a stress strain response of the thermoplastic polyester elastomer Hytrel 4556 \cite{Hytrel}, where nine stress-strain curves are given for temperatures between $\SI{-40}{\degreeCelsius}$ and $\SI{120}{\degreeCelsius}$. The temperature-dependent strain energy density function was determined for the best fit as follows: 
\begin{align}
\bar\Psi_{\text{Hytrel}} &= \ICdev \exp{\left(\sqrt{\left(- \bar T + \exp{\left(1.41 \sqrt{\left(0.5 \log{\bar T} - 1\right) \log{\bar T} }\right)} - 1.0\right) \log^{2}{\bar T}}\right)} \notag \\ &+ 0.131 \ICdev \log{\bar T } - 2 \IICdev \log{\bar T } + \exp{\left(41.19 \exp{\left(- \IICdev\right)}\right)} \log{\bar T }
\label{equ:HytrelBestFit}
\end{align}
\Rasulthree{where a scaled temperature $\bar T$, given as 
\begin{equation}
\bar T = \frac{T}{400} + \frac{1}{2}
\end{equation}
is utilized. Scaling the input variables to a similiar order of magnitude, while mathematically not necessary, is beneficial for the numerical performance of the optimization procedure.} The fitting was realized on a training set using only five temperature-dependent curves ($\SI{-40}{\degreeCelsius}$, $\SI{0}{\degreeCelsius}$, $\SI{40}{\degreeCelsius}$, $\SI{90}{\degreeCelsius}$, $\SI{120}{\degreeCelsius}$) and is visualized in \autoref{fig:TempFitGivenCurves}. The remaining four curves ($\SI{-20}{\degreeCelsius}$, $\SI{23}{\degreeCelsius}$, $\SI{60}{\degreeCelsius}$, $\SI{100}{\degreeCelsius}$) were not included during training, thus they can be used as a test set to evaluate the quality of the predicted strain energy function, as given in \autoref{fig:TempFitUnknownCurves}.

For both sets, even for the curves that have not been utilized during training, predicted stress response is very close to the experimental data. 

\section{Conclusion}\label{sec:conclusion}
In this article, we investigated the applicability of genetic programming and symbolic regression in the context of finding suitable strain energy functions. The necessary continuum mechanical framework and the principles of symbolic regression were presented. The chosen approach was evaluated using a benchmark test for the generalized Mooney-Rivlin model. With this method, one can successfully find the underlying strain energy function in a fairly short period of time. For verification the approach was tested for uniaxial tensile tests, equibiaxial tensile tests, and pure shear tests under large strains. In addition, a strain energy function for multi-axial loading of vulcanized rubber was determined. Subsequently, the method was applied to find a temperature-dependent strain energy function for a thermoplastic polyester elastomer. For both specified experimental data sets a very good fitting was obtained.

\Rasul{Compared to classical and data-driven approaches for finding material models, the method presented here is of numerous advantages. First of all, the determined strain energy function is provided as an algebraic equation. Thus, it is possible to interpret and discuss the resulting model from a physical point of view. This approach also provides a material model that can be used directly for the simulation of complex 3d components with arbitrary loads and boundary conditions. Secondly, the method is not subject to biases as only the experimental data is used to find the strain energy function. The fitting is solely based on the stress response from experimental data.} Thirdly, the approach can very easily be extended with further parameters. Temperature dependencies as well as process related parameters can be easily included into the proposed framework. Finally, the approach can be easily integrated into commercial software packages, since the derivatives of the strain energy density function with respect to the invariants are already calculated, making this method particularly interesting for practical applications.

\section*{Appendix}

\begin{table}[h!]
	\caption{Specified values for the {\tt SymbolicRegressor} class in the {\tt gplearn} package for all fits.}
	\label{tab:ParametersOverview}
	\centering\begin{tabular}{rccccc}
		\toprule
		Inputs& gMR case 1 & gMR case 2 &  gMR case 3 & Treloar & Hytrel \\
		\midrule
		{\tt population\_size} & 1000 & 7500 & 15000 & 5000 & 15000 \\
		{\tt generations} & 50 & 50 & 50 & 200 & 50 \\
		{\tt stopping\_criteria}& 0.001 & 0.001 & 0.001 & 0.001  &  0.001 \\
		{\tt p\_crossover}& 0.7 & 0.7 & 0.6 & 0.65 & 0.6 \\
		{\tt p\_subtree\_mutation} & 0.15 & 0.15 & 0.15 & 0.1 & 0.15 \\
		{\tt p\_hoist\_mutation} & 0.1 & 0.1 & 0.1 & 0.05 & 0.1 \\
		{\tt p\_point\_mutation} & 0.05 & 0.05 & 0.15 & 0.1 & 0.15 \\
		{\tt max\_samples} & 0.9 & 0.9 & 0.9  & 0.6 & 0.9 \\
		{\tt parsimony\_coefficient} & 0.003 & 0.003 & 0.0005 & 0.0025 & 0.02 \\
		\bottomrule
	\end{tabular}
\end{table}


\bibliography{Mendeley,Buecher}
\end{document}